\documentclass{article}
\usepackage{spconf,amsmath,graphicx}
\usepackage[hidelinks]{hyperref}

\title{Perfect Scores, Misplaced Explanations: A Controlled Study of Explainability in Audio Deepfake Detection}

\name{David Combei,$^1$ Dan Oneata,$^{2,3}$ Francesco Paissan,$^{4,5}$ Cem Subakan,$^{4,5}$ Horia Cucu,$^2$ Adriana Stan$^1$\thanks{Corresponding author: \texttt{adriana.stan@com.utcluj.ro}}}
\address{$^1$Technical University of Cluj-Napoca, Romania\quad $^2$POLITEHNICA Bucharest, Romania \\
$^3$Bitdefender, Romania \quad
$^4$Mila-Quebec AI Institute, Canada \quad
$^5$Universit\'e Laval, Canada}

\newcommand{\rn}[1]{\textcolor{gray}{\scriptsize #1}}  

\usepackage{comment}
\usepackage[textsize=tiny]{todonotes}
\usepackage{microtype}
\usepackage{booktabs}
\usepackage{tabularx}
\usepackage{stmaryrd}
\usepackage{xspace}

\newcommand{\ab}{\mathbf{a}}
\newcommand{\mb}{\mathbf{m}}
\newcommand{\spec}{\mathbf{s}}

\definecolor{myblue}{HTML}{1f78b4}
\definecolor{mybluelight}{HTML}{a6cee3}

\definecolor{mygreen}{HTML}{fb9a99}
\definecolor{mygreenlight}{HTML}{33a02c}

\definecolor{mypurple}{HTML}{6a3d9a}
\definecolor{mypurplelight}{HTML}{cab2d6}

\definecolor{myred}{HTML}{fdbf6f}
\definecolor{myredlight}{HTML}{e31a1c}

\definecolor{myolive}{HTML}{ada136}
\definecolor{myolivelight}{HTML}{7ead36}

\begin{document}
%
\maketitle
\begin{abstract}
What if an explanation achieves perfect scores but it is uninterpretable? We show that this happens when explaining audio deepfake detectors. Using an optimization-based method, we explain a state-of-the-art wav2vec2-based detector. The explanations achieve optimal scores on standard explainability metrics, i.e. fidelity and faithfulness. Visual inspection, however, reveals unintuitive and uninterpretable patterns. 
To understand this discrepancy, we run a controlled perturbation study in which we modify known frequency regions of the speech signal. Although we propose a model that is able to obtain close-to-optimal results with respect to the metrics, the explanations locate the manipulated region only when it is fixed across samples, and mostly fall to chance level when it varies. We further show that the signal used to fill the removed regions largely determines the explanation and its localization, without being reflected in the metrics.


\end{abstract}
%

\section{Introduction}

What makes this speech sample a \emph{fake} one? A state-of-the-art deepfake detector can accurately tell that a file \emph{is} generated, but not \emph{what} led to that decision, and accuracy alone does not translate into trust~\cite{JainDE22}. Understanding these decisions is necessary in high-risk setups such as political disinformation or digital evidence integrity.
Explainable AI (xAI) aims to reduce this opaqueness. Explainable-by-design approaches~\cite{rudin19} are hard to build, as deepfake detection is difficult even for expert human observers~\cite{MullerPW22}. Interpretable classifiers trained on human-understandable features, e.g. breathing patterns~\cite{LaytonAOWGBT25}, fundamental frequency~\cite{xue22f0} or handcrafted acoustic features~\cite{PascuOCM25}, still lag behind high performing detectors.
Such detectors are built on large self-supervised learning (SSL) models, in particular the wav2vec2 family~\cite{wav2vec2}, and have been shown to generalize well across datasets and manipulation types~\cite{ ge2025asru,oneata_is24}. 
Linear classifiers~\cite{ge2025asru,combei_interspeech25}, 
graph neural networks~\cite{tak2022odyssey} and state space models~\cite{tran2025leveraging} are generally trained on top of these SSL features. But the added complexity makes their decisions harder to interpret.
Explainable audio deepfake detection has so far relied on gradient-based explanations in the time domain~\cite{grinberg2025does}, on predicting spectrogram differences between real and vocoded signals~\cite{grinberg2025data}, on artifact-specific expert detectors~\cite{negroni2026interpretable}, or on one-class anomaly detection~\cite{coletta2025anomaly}.

\begin{figure}[t!]
    \centering
    \includegraphics[width=.9\linewidth]{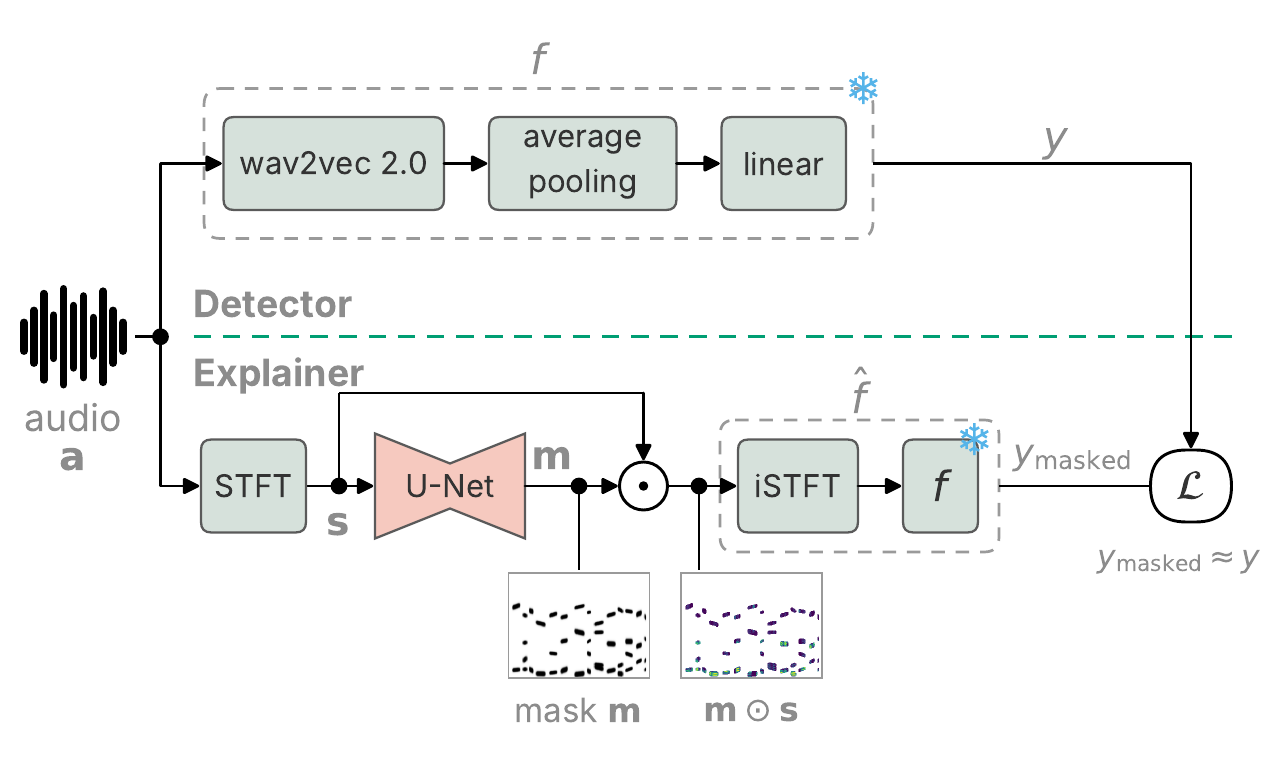}
    \caption{%
        Proposed xAI method. For a frozen audio deepfake detector $f$, we learn an explainer that predicts soft, sparse masks $\mb$ over the input spectrogram \textbf{s}. Keeping only the masked regions should preserve the detector’s prediction, while removing them should flip it.
    }
    \label{fig:method}
\end{figure}

We aim to explain a complex, high-performing wav2vec2-based deepfake detector by pinpointing the time--frequency regions relevant to its decision, i.e. \emph{saliency maps}. Our explanations are soft masks over the input spectrogram, obtained by maximizing two standard explainability metrics: \emph{fidelity}, which checks that the prediction is preserved when only the masked regions are kept, and \emph{faithfulness}, which checks that the prediction changes when the regions are removed. The metrics are optimized across many training samples, yielding an amortized optimization-based method~\cite{chen2018learning}. The work is similar to methods for sound~\cite{paissan2024lmac}, image~\cite{fong2017iccv} and text~\cite{hase2021out} classification, but it is applied for the first time to audio deepfake detection.

We find that our method achieves near-perfect scores on both xAI metrics, surpassing gradient-based explanation methods, which inherently struggle to achieve high faithfulness. 
To assess whether these metrics produce meaningful explanations, we conduct a controlled evaluation in which specific frequency bands are perturbed by replacing them with their vocoded counterparts.
While the deepfake detector is able to unveil such manipulations,
none of the gradient-based methods recovers the true manipulated regions. Our explainer locates the band when it is the same across samples, but when it varies, localization mostly drops to chance.
Finally, the signal used to fill the removed regions acts as the baseline of the explanation~\cite{sturmfels2020visualizing}, and changing it moves localization from well above chance to chance level, without impacting the metrics.

To summarize, our contributions are:
\textbf{(i)}~we build a strong amortized, optimization-based explainer that produces time–frequency masks for a wav2vec2-based deepfake detector;
\textbf{(ii)}~we show that the model attains optimal scores on standard explainability metrics, but fails to provide interpretable explanations;
\textbf{(iii)}~using controlled frequency perturbations we show that high fidelity and faithfulness do not automatically imply correct localization;
\textbf{(iv)}~we show that the fill of the masked-out regions determines the explanations and their localization, but metrics do not reflect it.

\vspace{-.2cm}
\section{Proposed xAI method}

Our method pairs a frozen detector with a trainable explainer. The explainer predicts a mask highlighting the spectral regions most relevant to the detector's prediction (see Figure~\ref{fig:method}).


\vspace{-.2cm}
\subsection{The detector}
The detector is a binary classifier $f$ that takes a 1D time-domain audio waveform $\ab$ as input and outputs one of two labels $y \in \left\{\mathrm{real},\mathrm{fake}\right\}$.
It comprises: a wav2vec 2.0 feature extractor \cite{wav2vec2}, a global average pooling layer, and a linear layer for classification.
We use the wav2vec2 XLS-R 2B backbone, with features extracted from layer 9 out of 48---early layers were found to work better~\cite{elkheir2025comprehensive}.
Audio is sampled at 16 kHz. 
The classifier head is a \texttt{scikit-learn} logistic regression with low regularization ($C=10^6$).

\vspace{-.2cm}
\subsection{The explainer}
Since the detector operates on waveforms, explanations computed directly on its input would only reveal \textit{when} information is important, whereas most speech experts would expect explanations in the time$\times$frequency domain. We therefore train an \emph{explainer} that generates a soft saliency mask $\mb$ over the spectrogram $\spec$ of the input audio.
Values near 1 mark evidence for the prediction $y$ and values near 0 mark irrelevant regions. 
The masked spectrogram is converted back with the differentiable iSTFT---enabling the explanations to be obtained by either optimization or gradients--- and fed to the frozen detector, yielding a wrapped detector
$\hat f(\mathbf{s}) = f(\mathrm{iSTFT}(\mathbf{s}))$. The wrapped detector returns the probability that the input is fake. STFT uses the same parameters as HiFi-GAN: a window length of 1024 samples, FFT size of 1024, and hop length of 256.
The explainer is trained across many samples 
which makes inference faster and less sensitive to spurious patterns in individual inputs~\cite{dabkowski2017neurips}.
The architecture of the explainer uses a U-Net with three 2D convolutional layers
and an output sigmoid activation which constrains the output explanation to the $[0, 1]$ interval.
Unlike~\cite{jethani2021have}, we keep the classifier frozen, which prevents the explainer from encoding label information. And although the output is linear, the wav2vec 2.0 front-end makes the decision boundary highly non-linear.

The \textbf{explanation} is a mask $\mathbf{m}\in[0,1]^{F\times T}$ of the same size
as the spectrogram $\mathbf{s}$. Regions removed by the mask are replaced by a fill signal $\mathbf{z}$. We denote the input that keeps only
the masked-in regions ($\mathbf{s}_{\mathbf{m}}$), and the input that retains the masked-out regions ($\mathbf{s}_{\bar{\mathbf{m}}} $), as:
\begin{align}
  \mathbf{s}_{\mathbf{m}} &= \mathbf{m}\odot\mathbf{s} + (1-\mathbf{m})\odot\mathbf{z}, \\
  \mathbf{s}_{\bar{\mathbf{m}}} &= (1-\mathbf{m})\odot\mathbf{s} + \mathbf{m}\odot\mathbf{z},
\end{align}
where $\odot$ denotes element-wise multiplication.


The optimization uses the following terms:

(a) \textbf{fidelity} checks that keeping only the masked regions preserves the prediction:
\begin{equation}
  \mathrm{FID} := \big[\!\big[\, H(\hat f(\mathbf{s})) = H(\hat f(\mathbf{s}_{\mathbf{m}})) \,\big]\!\big],
\end{equation}
where $H$ thresholds the probability at 0.5.

(b) \textbf{faithfulness} measures the drop in probability when the masked regions are removed:
\begin{equation}
  \mathrm{FTH} := \hat f(\mathbf{s}) - \hat f(\mathbf{s}_{\bar{\mathbf{m}}}).
\end{equation}

(c) \textbf{sparsity}: the mask should keep as little of the spectrogram as
possible, $\|\mathbf{m}\|_1$,

\noindent which are combined in the training loss:
\begin{align}
  \mathcal{L} = \alpha\,\mathrm{BCE}\big(\hat f(\mathbf{s}_{\mathbf{m}}),\, y\big) 
              + \beta\,\mathrm{BCE}\big(\hat f(\mathbf{s}_{\bar{\mathbf{m}}}),\, 1-y\big) 
              + \gamma\,\|\mathbf{m}\|_1 ,
\end{align}
where $y$ is the detector's prediction on the original input, BCE is the binary cross-entropy,  and $\left\{\alpha, \beta, \gamma \right\}$ 
are the weights.

The \textbf{fill ($\mathbf{z}$)} acts as the null point of the explanation: the mask is learned to decide which regions to keep and which to replace by $\mathbf{z}$, so the choice of $\mathbf{z}$ defines what the explanation is relative to. We compare four fills:
(a)~the \emph{zero fill}, where $\mathbf{z}$ is all zeros;
(b)~the \emph{average fill}, where $\mathbf{z}$ is the average spectrogram over the training data;
(c)~the \emph{real fill}, where $\mathbf{z}$ is the spectrogram of a paired real recording---same speaker, same content, but no temporal alignment;
(d)~the \emph{optimized fill}, a single spectrogram $\mathbf{z}$ learned such that the detector is maximally uncertain on it, $\hat f(\mathbf{z}) = 0.5$, while preserving the prediction when combined with real or fake inputs under various masks. The optimized fill is bootstrapped: it is re-learned using masks generated by the trained explainer, and the explainer is then re-trained with the new fill.

\begin{table*}[h!]
    \small
    \caption{Fidelity (FID), faithfulness (FTH), localization AUC and relevance mass accuracy (RMA) on realistic data (MLAAD) and on two controlled setups where a fixed or a varying 1~kHz band is vocoded. Rows 0--4 use the optimized fill to compute the scores. Chance: AUC 0.5, RMA 0.125.}
    \label{tab:main-results}
    \centering
    \setlength{\tabcolsep}{5pt}
    \begin{tabular}{@{}rl cc c cccc c cccc@{}}
        \toprule
            &  & \multicolumn{2}{c}{\textbf{Realistic} {\scriptsize(EER 7.4\%)}} & & \multicolumn{4}{c}{\textbf{Fixed band} {\scriptsize(EER 0.0\%)}} & & \multicolumn{4}{c}{\textbf{Varying band} {\scriptsize(EER 0.04\%)}} \\
        \cmidrule{3-4} \cmidrule{6-9} \cmidrule{11-14}
            & Method & FID$\uparrow$ & FTH$\uparrow$ & & FID$\uparrow$ & FTH$\uparrow$ & AUC$\uparrow$ & RMA$\uparrow$ & & FID$\uparrow$ & FTH$\uparrow$ & AUC$\uparrow$ & RMA$\uparrow$ \\
        \midrule
        \rn{1} & Random          & 0.993 & 0.005 & & \bf 1.000 & 0.000 & 0.500 & 0.125 & & 0.981 & 0.020 & 0.500 & 0.125 \\
        \rn{2} & Saliency        & 0.855 & 0.057 & & 0.824 & 0.000 & 0.497 & 0.090 & & 0.999 & 0.004 & 0.496 & 0.118 \\
        \rn{3} & Inp$\times$Grad & 0.975 & 0.085 & & 0.755 & 0.003 & 0.516 & 0.137 & & 0.328 & 0.001 & 0.509 & 0.101 \\
        \midrule
        \rn{4} & \textbf{Ours}, optimized fill & \bf 1.000 & \bf 0.999 & & \bf 1.000 & \bf 1.000 & 0.811 & 0.289 & & \bf 1.000 & \bf 1.000 & 0.505 & 0.124 \\
        \rn{5} & \textbf{Ours}, real fill & 0.931 & 0.886 & & \bf 1.000 & 0.999 & \bf 0.897 & \bf 0.436 & & \bf 1.000 & \bf 1.000 & \bf 0.728 & \bf 0.358 \\
        \rn{6} & \textbf{Ours}, average fill & \bf 1.000 & \bf 0.999 & & \bf 1.000 & 0.986 & 0.448 & 0.136 & & \bf 1.000 & 0.997 & 0.495 & 0.128 \\
        \rn{7} & \textbf{Ours}, zero fill & \bf 1.000 & \bf 0.999 & & \bf 1.000 & 0.994 & 0.713 & 0.409 & & \bf 1.000 & 0.999 & 0.498 & 0.126 \\
        \bottomrule
    \end{tabular}
\end{table*}

\vspace{-0.2cm}
\subsection{The baselines}
We compare against two widely used, training-free gradient-based methods, applied to the same wrapped detector $\hat f$ via Captum: saliency~\cite{simonyan2014deep} and input $\times$ gradient~\cite{shrikumar2016not}.
We also use a random baseline that assigns uniform scores in $[0,1]$ to spectrogram superpixels~\cite{felzenszwalb2004efficient} to test whether the metrics can distinguish meaningful explanations from arbitrary ones. A superpixel is a patch of neighboring time-frequency bins that belong together visually, e.g. part of a harmonic or silence.

\subsection{The results}

\textbf{Data.}
We subsample 10\% of MLAADv9~\cite{mlaad} by selecting examples for each of the 298 model–language combinations, amounting to 27k fake samples.
27k real samples are selected from M-AILABS~\cite{mailabs}.
The data is split 80:20 into train and test sets for the deepfake detector. The explainer is trained and evaluated only on fakes. This is because a fake decision likely relies on a localized anomaly, such as an altered frequency band, whereas a real decision may rely on the absence of such anomalies, which a mask cannot highlight.
In all setups the speech files are cropped or zero-padded to 5 seconds.

\textbf{Quantitative results.}
The detector reaches 7.4\% equal error rate (EER) meaning that it can mostly discriminate fake from real samples. Table~\ref{tab:main-results}, column Realistic, shows the fidelity and faithfulness results for all xAI methods.
Saliency and input $\times$ gradient obtain good fidelity, but their faithfulness is close to 0 (rows 2--3). Even the random baseline (row~1) obtains a fidelity of 0.99, although its explanations are arbitrary.
Meanwhile, our explainer (rows 4--7) achieves high fidelity and faithfulness irrespective of the fill signal.  

\begin{figure*}[t]
    \centering
    \includegraphics[width=.75\linewidth]{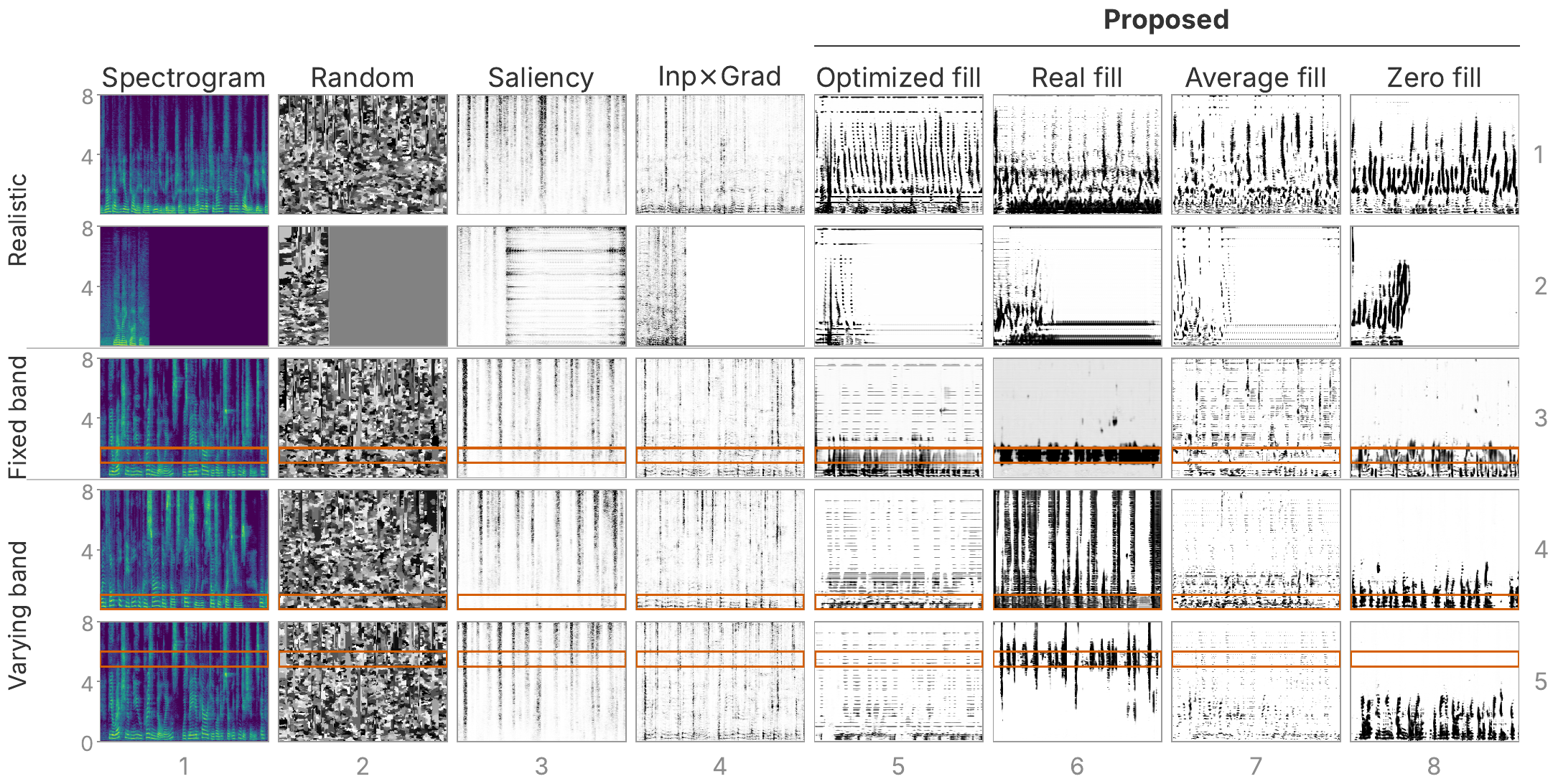}
    \caption{%
        Explanations for two MLAAD fake samples (rows 1--2), a fixed-band sample (1--2~kHz, row 3) and two varying-band samples (0--1~kHz and 5--6~kHz, rows 4--5). Columns 5--8 show our masks for each fill; the baselines' maps do not depend on the fill. The orange box marks the manipulated band. Each mask is normalized separately.
    }
    \label{fig:qualitative}
\end{figure*}

\textbf{Qualitative results.}
But how interpretable are the produced explanations? Figure~\ref{fig:qualitative}
shows explanations for two MLAAD samples (rows 1--2). The random baseline (column 2) treats the padded silence in row 2 as a single patch, yet obtains near-perfect fidelity. Saliency (column 3) highlights broad vertical bands and also assigns relevance to the padded silence. Input$\times$Gradient (column 4) follows fine structure such as onsets and low-frequency harmonics, and vanishes in silence.
In contrast, our masks (columns 5--8) are sharper. Yet the four fills, which obtain equally high scores, produce very different masks for the same input, e.g. only some of them extend into the padded silence of row 2. Such patterns are difficult to relate to any plausible synthesis artifact, and point to the fact that high xAI scores do not necessarily translate into identifiable manipulation cues in audio. We examine this discrepancy with controlled experiments in the next section.


\section{Understanding the explanations}

To test whether the explainer can truly locate synthesis artifacts---albeit not human-interpretable---we evaluate it on speech where known spectral regions have been manipulated.
We start with 5k real audio samples from LJ Speech (single speaker, high quality)~\cite{ljspeech17} and resample them to 16~kHz.
Each sample is vocoded with a HiFi-GAN vocoder~\cite{kong2020hifi} trained on LibriTTS \cite{zen2019libritts}.
We then replace a single 1~kHz band (out of 8 possible) with the corresponding band from its time-aligned vocoded version. The replacement is performed in the complex spectrum domain. The replaced band is either the same across all samples, or varies from sample to sample.
We first train the deepfake detector on the newly generated datasets. Both setups achieve low EERs of 0.0\% (fixed band) and 0.04\% (varying band).
Table~\ref{tab:main-results} also includes the metrics for these setups. 
As in the realistic case, the gradient-based methods have low faithfulness, while our explainer yields near-perfect fidelity and faithfulness.

\textbf{Localization and the effect of the fill.}
Since the manipulated regions are known, we can measure their localization with the area under the ROC curve (AUC), using the mask values as scores and the manipulated bins as positives.
We can also report the relevance mass accuracy (RMA)~\cite{arras2022clevr}, the fraction of the total soft mask value that falls inside the manipulated band. It equals 1 when all relevance lies in the band, and 0.125 for a random explanation, i.e. the band’s share of the spectrogram (1 of 8~kHz).

All gradient-based methods remain at chance level (AUC $\approx$ 0.50, RMA $\approx$ 0.125).
With the optimized fill, our explainer localizes the fixed band (AUC 0.81) but places less than a third of its relevance inside it (RMA 0.29). Its mask covers the band together with most of the spectral region below it (Figure~\ref{fig:qualitative}, row 3, column 5). When the band varies, localization drops to chance (AUC 0.51, RMA 0.12), and the masks show the same patterns regardless of the manipulation (rows 4--5, column 5). 

A surprising result is the fact that
the explainer's ability to localize the manipulation largely depends on the fill (rows 4--7 of Table~\ref{tab:main-results}).
The real fill localizes best (AUC 0.90 and 0.73), and its masks closely follow the manipulated band (rows 3 and 5, column 6 of Figure~\ref{fig:qualitative}). It is the only fill whose mask moves to the 5--6~kHz band in row 5.
However, the real and manipulated signals differ only in the swapped band, so the fill itself conveys the location of the manipulation during training.
The average fill yields scattered masks and chance-level localization (column 7). The zero fill (column 8) behaves like the optimized one: it covers the fixed band together with the region below it, stays at low frequencies when the band varies, and leaves the 5--6~kHz band empty in row 5.

This behaviour may be partly explained by passing each fill on its own through the detectors. 
The zero fill is classified as fake by all three detectors; the average fill is classified as fake by the two controlled ones, while the realistic detector classifies it as real. With these fills, a sparse mask of a fake sample stays fake wherever it lies. 
Fidelity remains high even when the fill is not biased. The optimized fill is neutral-by-design on its own, and the real fill pushes the prediction towards real, yet arbitrary masks still preserve the prediction: a random explanation obtains a fidelity of 0.98--1.00 with the optimized fill and 0.98 with the real fill on MLAAD (not shown in the table), compared with 0.93 for our explainer. Since soft masks keep part of the input everywhere, enough of the fake signal survives for the detector to keep its decision, so fidelity cannot separate meaningful explanations from arbitrary ones.
On the other hand, alternating the training of the optimized fill with the explainer's may let the two co-adapt~\cite{jethani2021have}. But the zero fill, which is not learned, yields the same fixed masks (rows 4--5, columns 5 and 8), so co-adaptation alone does not explain them. Nor is it a capacity limit, as the explainer architecture is able to locate randomly placed rectangles at AUC 1.0.

Put simply, the metrics alone cannot tell whether an explanation points to the real manipulation. And in our setup, the fill appears to decide much of what both the explanation and its scores look like.

\vspace{-.3cm}
\section{Conclusions}

We introduced an optimization-based method that explains a wav2vec2-based deepfake detector with time--frequency masks. It reaches near-perfect fidelity and faithfulness, yet controlled manipulations with a known location show that its explanations often miss the manipulated regions. With the optimized and zero fills, the explainer localized a band fixed across samples, but not one whose position varied. The real fill localized both, but only because its training process reveals the band. Meanwhile, the metrics stayed near-perfect, and even random explanations obtained near-perfect fidelity. These metrics should therefore not be used on their own to validate explanations and should be complemented by controlled manipulations with known ground truth.

These results may seem expected, as the method optimizes the very metrics we question (Goodhart's law: an optimized metric stops being a reliable measure). However, the problem extends beyond our method: these metrics can be inflated deliberately~\cite{hsia2024goodhart}, and we show that it also happens unintentionally. Designing more robust metrics is left for future work.

\clearpage
\textbf{Acknowledgements.} Cem Subakan is supported by the NSERC project RGPIN-2023-05759. Francesco Paissan is funded by Fonds de recherche du Québec (FRQ) and by the Institut intelligence et données (IID) at Université Laval

\textbf{Compliance with ethical standards.} This study was conducted retrospectively using human subject data made available in open access (MLAAD, M-AILABS and LJ Speech). Ethical approval was not required, as confirmed by the licenses attached to these datasets. 

\textbf{LLM use}. The authors used Claude (Anthropic) to edit and shorten the text, to write the plotting scripts for Figures 1 and 2, and to implement the localization metrics (AUC and RMA). All experiments, results and conclusions are the authors' own; all AI-assisted text and code were reviewed and verified by the authors, who take full responsibility for the content.

\bibliographystyle{IEEEbib}
\small
\bibliography{refs-short}

@article{sturmfels2020visualizing,
  author  = {Pascal Sturmfels and Scott Lundberg and Su-In Lee},
  title   = {Visualizing the Impact of Feature Attribution Baselines},
  journal = {Distill},
  year    = {2020},
  doi     = {10.23915/distill.00022}
}

@inproceedings{mlaad,
  title={{MLAAD}: The Multi-Language Audio Anti-Spoofing Dataset},
  author={N. M. Muller and P. Kawa and W. H. Choong and others},
  booktitle={IJCNN},
  year={2024},
}

@inproceedings{wav2vec2,
 author = {Baevski, A. and Zhou, Y. and Mohamed, A. and others},
 booktitle = {NeurIPS},
 title = {Wav2vec 2.0: A Framework for Self-Supervised Learning of Speech Representations},
 year = {2020}
}

@inproceedings{tak2022odyssey,
  author       = {H. Tak and M. Todisco and X. Wang and others},
  title        = {Automatic Speaker Verification Spoofing and Deepfake Detection Using
                  Wav2vec 2.0 and Data Augmentation},
  booktitle    = {Odyssey},
  year         = {2022},
}

@inproceedings{kong2020hifi,
  title={Hi{F}i-{GAN}: Generative adversarial networks for efficient and high fidelity speech synthesis},
  author={Kong, J. and Kim, J. and Bae, J.},
  booktitle={NeurIPS},
  year={2020}
}

@article{scikit-learn,
  title={Scikit-learn: Machine Learning in {P}ython},
  author={Pedregosa, F. and Varoquaux, G. and Gramfort, A. and others},
  journal={Journal of Machine Learning Research},
  volume={12},
  year={2011}
}

@inproceedings{grinberg2025data,
  title={A Data-Driven Diffusion-based Approach for Audio Deepfake Explanations},
  author={Grinberg, P. and Kumar, A. and Koppisetti, S. and others},
  booktitle={Interspeech},
  year={2025}
}

@inproceedings{paissan2024lmac,
  title={Listenable maps for audio classifiers},
  author={Paissan, F. and Ravanelli, M. and Subakan, C.},
  booktitle={ICML},
  year={2024}
}

@inproceedings{dabkowski2017neurips,
  title={Real time image saliency for black box classifiers},
  author={Dabkowski, P. and Gal, Y.},
  booktitle={NeurIPS},
  year={2017}
}

@inproceedings{fong2017iccv,
  title={Interpretable explanations of black boxes by meaningful perturbation},
  author={Fong, R. C. and Vedaldi, A.},
  booktitle={ICCV},
  year={2017}
}

@inproceedings{ge2025asru,
  title={Post-training for deepfake speech detection},
  author={Ge, W. and Wang, X. and Liu, X. and others},
  booktitle={ASRU},
  year={2025}
}

@misc{mailabs,
   author = {Solak, I. C. A. and Naumov, D.},
   title = {The {M-AILABS} Speech Dataset},
   year = {2017},
   howpublished = {\url{https://github.com/i-celeste-aurora/m-ailabs-dataset}},
   note         = {Accessed on 05/02/2026}
}

@misc{ljspeech17,
  author       = {K. Ito and L. Johnson},
  title        = {The {LJ} {S}peech Dataset},
  howpublished = {\url{https://keithito.com/LJ-Speech-Dataset/}},
  year         = 2017,
 
}

@INPROCEEDINGS{combei_interspeech25,
  author={Combei, D. and Stan, A. and Oneata, D. and others},
  booktitle={Interspeech},
  title={Unmasking real-world audio deepfakes: A data-centric approach},
  year={2025},
}

@inproceedings{oneata_is24,
  author = {Pascu, O. and Stan, A. and Oneata, D. and others},
  title = {Towards generalisable and calibrated audio deepfake detection with self-supervised representations},
  booktitle= {Interspeech},
  year = 2024,
}

@inproceedings{zen2019libritts,
  title={Libri{TTS}: A corpus derived from librispeech for text-to-speech},
  author={Zen, H. and Dang, V. and Clark, R. and others},
  booktitle={Interspeech},
  year={2019}
}

@inproceedings{MullerPW22,
  author = {N. M. Müller and K. Pizzi and J. Williams},
  title = {Human Perception of Audio Deepfakes},
  year = {2022},
  booktitle = {DDAM@MM},
}

@article{JainDE22,
  author = {A. K. Jain and D. Deb and J. J. Engelsma},
  title = {Biometrics: Trust, But Verify},
  year = {2022},
  journal = {IEEE Trans. Biom. Behav. Identity Sci.},
}

@article{LaytonAOWGBT25,
  author = {S. Layton and T. D. Andrade and D. Olszewski and others},
  title = {Every Breath You Don't Take: Deepfake Speech Detection Using Breath},
  year = {2025},
  journal = {Digit. Threat. Res. Pract.},
}

@inproceedings{PascuOCM25,
  author = {O. Pascu and D. Oneata and H. Cucu and others},
  title = {Easy, Interpretable, Effective: open{SMILE} for voice deepfake detection.},
  year = {2025},
  booktitle = {ICASSP},
}

@inproceedings{xue22f0,
  author       = {J. Xue and C. Fan and Z. Lv and others}, title        = {Audio Deepfake Detection Based on a Combination of {F0} Information and Real Plus Imaginary Spectrogram Features},
  booktitle    = {DDAM@MM},
  year         = {2022},
}

@article{rudin19,
  author       = {C. Rudin},
  title        = {Stop explaining black box machine learning models for high stakes decisions and use interpretable models instead},
  journal      = {Nat. Mach. Intell.},
  year         = {2019},
}

@inproceedings{tran2025leveraging,
  title={Leveraging {SSL} Speech Features and {M}amba for Enhanced DeepFake Detection},
  author={Tran, H. M. and Lolive, D. and Guennec, D. and others},
  booktitle={Interspeech},
  year={2025}
}

@inproceedings{negroni2026interpretable,
  author    = {Negroni, V. and Wang, X. and Ge, W. and others},
  title     = {Toward Interpretable Speech Deepfake Detection using Artifact-Specific Experts and Calibrated Detection Scores},
  booktitle = {ACM Deepfake Forensics Workshop}, 
  year      = {2026},
  doi       = {10.1145/3840473.3840518}
}

@article{arras2022clevr,
  title   = {{CLEVR-XAI}: A benchmark dataset for the ground truth evaluation of neural network explanations},
  author  = {Arras, L. and Osman, A. and Samek, W.},
  journal = {Inf. Fusion},
  year    = {2022},
}

@inproceedings{coletta2025anomaly,
  author    = {Coletta, E. and Salvi, D. and Negroni, V. and others},
  title     = {Anomaly Detection and Localization for Speech Deepfakes via Feature Pyramid Matching},
  booktitle = {EUSIPCO},
  year      = {2025}
}

@inproceedings{hsia2024goodhart,
  title={Goodhart’s law applies to {NLP}'s explanation benchmarks},
  author={Hsia, J. and Pruthi, D. and Singh, A. and others},
  booktitle={Findings EACL},
  year={2024}
}

@inproceedings{hase2021out,
  title={The out-of-distribution problem in explainability and search methods for feature importance explanations},
  author={Hase, P. and Xie, H. and Bansal, M.},
  booktitle={NeurIPS},
  year={2021}
}

@inproceedings{chen2018learning,
  title={Learning to explain: An information-theoretic perspective on model interpretation},
  author={Chen, J. and Song, L. and Wainwright, M. and others},
  booktitle={ICML},
  year={2018},
}

@inproceedings{jethani2021have,
  title={Have we learned to explain? {H}ow interpretability methods can learn to encode predictions in their interpretations},
  author={Jethani, N. and Sudarshan, M. and others},
  booktitle={AISTATS},
  year={2021},
}

@inproceedings{simonyan2014deep,
  title={Deep inside convolutional networks: Visualising image classification models and saliency maps},
  author={Simonyan, K. and Vedaldi, A. and Zisserman, A.},
  booktitle={ICLR Workshop},
  year={2014}
}

@article{shrikumar2016not,
  title={Not just a black box: Learning important features through propagating activation differences},
  author={Shrikumar, A. and Greenside, P. and Shcherbina, A. and others},
  journal={arXiv preprint arXiv:1605.01713},
  year={2016}
}

@inproceedings{elkheir2025comprehensive,
  title={Comprehensive layer-wise analysis of {SSL} models for audio deepfake detection},
  author={El Kheir, Y. and Samih, Y. and Maharjan, S. and others},
  booktitle={Findings NAACL},
  year={2025}
}

@inproceedings{grinberg2025does,
  title={What does an audio deepfake detector focus on? {A} study in the time domain},
  author={Grinberg, P. and Kumar, A. and Koppisetti, S. and others},
  booktitle={ICASSP},
  year={2025},
}

@article{felzenszwalb2004efficient,
  title={Efficient graph-based image segmentation},
  author={Felzenszwalb, P. and Huttenlocher, D.},
  journal={Int. J. Comput. Vis.},
  volume={59},
  number={2},
  year={2004},
}

\end{document}